%% file: paper.tex
\documentclass[]{spie}  

\usepackage{euclid}
\usepackage[square,sort&compress,numbers]{natbib}
\usepackage{amsmath,amsfonts,amssymb}
\usepackage{graphicx}
\usepackage[colorlinks=true, allcolors=blue]{hyperref}
\usepackage[nolist,nohyperlinks]{acronym}
\usepackage[labelformat=simple]{subcaption}
\usepackage[normalem]{ulem}
\usepackage{multirow}
\usepackage{booktabs}
\usepackage{rotating}
\usepackage{orcidlink}
\usepackage{endnotes}
\usepackage{siunitx} 
\title{Performance of the image persistence model for {\LARGE\textit{\textbf{Euclid}}} infrared detectors}

\input{authors}

\begin{document} 
\maketitle

\begin{abstract}
Large-format infrared detectors are at the heart of major ground and space-based astronomical instruments, and the HgCdTe HxRG is the most widely used. The Near Infrared Spectrometer and Photometer (NISP) of the ESA's \Euclid mission launched in July 2023 hosts 16 H2RG detectors in the focal plane. Their performance relies heavily on the effect of image persistence, which results in residual images that can remain in the detector for a long time contaminating any subsequent observations. 
Deriving a precise model of image persistence is challenging due to the sensitivity of this effect to observation history going back hours or even days. Nevertheless, persistence removal is a critical part of image processing because it limits the accuracy of the derived cosmological parameters.
We will present the empirical model of image persistence derived from ground characterization data, adapted to the \Euclid observation sequence and compared with the data obtained during the in-orbit calibrations of the satellite. 
  
\end{abstract}


\section{INTRODUCTION}
\label{sec:intro}

\Euclid is an M2 mission of the European Space Agency (ESA)'s Cosmic Vision 2015 -- 2025 program \citep{Euclid_overview}. It is a cosmological survey designed to explore the origins of the accelerated expansion of our Universe and to probe the nature of dark energy (DE) and dark matter (DM) which together contribute to about 95\% of the mass-energy content of the Universe and which play a fundamental role in the formation and evolution of large-scale structures.  Two main cosmological probes are used: galaxy clustering requiring accurate measurements of galaxy redshifts and weak gravitational lensing using precise measurements of galaxy shape distortions by DM. As a result, a six-year survey covering 14\,000 deg$^2$ of the sky will provide a 3D map of about 1.5 billion galaxies up to redshift $z\sim2$.

The \Euclid spacecraft is equipped with two instruments: a Visible Imager (VIS) designed for high-resolution imaging with an angular resolution of 0.1 arcsec pixel$^{-1}$ \citep{VIS_overview} and the Near Infrared Spectrometer and Photometer (NISP) \citep{NISP_overview} dedicated to redshift measurements. 

The NISP focal plane consists of 16 HgCdTe-based H2RG detectors (Teledyne) arranged in an array of $4 \times 4$ in the Focal Plane Assembly (FPA). The main properties of the sensors were extensively characterised during ground characterisation campaigns (\citep{2015SPIE.9602E..0GS}, \citep{2016SPIE.9915E..1YS}, \citep{perf_h2rg}, \citep{barbier18}). While the noise properties of the NISP detectors satisfy the requirements, image persistence is one of the most problematic effects. During \Euclid nominal observations the NISP works in photometric mode using three filters: \YE (950--1212nm), \JE (1168--1567nm) and \HE (1522--2021nm) mounted on a filter wheel, and in spectrometric mode using grisms (1206--1892nm or 926--1366nm) mounted on a grism wheel and dispersing in either $0^\circ$, $180^\circ$ or $270^\circ$ direction (\citep{Schirmer-EP18}, \citep{NISP_flight}). NISP continuously scans the sky using a reference observation sequence (ROS) of consecutive pointings. Each pointing consists of four dithers and each dither includes one spectrometric exposure of about 550s followed by three photometric exposures of about 87s. Between consecutive exposures, from a dozen to a few tens of seconds are dedicated to the wheels movement, the telescope pointing, and the stabilization of the system. Thus, bright objects observed in one exposure leave persistence traces in subsequent exposures and the effects can accumulate over the time of sky scanning. Detecting and masking or correcting these post-images is therefore of paramount importance for the scientific outcome of the \Euclid mission since persistence can significantly affect the accuracy of redshift measurements.

The widely accepted phenomenological model of physics behind the effects of image persistence was proposed by Smith (\citep{Smith_2008a}, \citep{Smith_2008b}). This model assumes that persistence is due to charge trapping and release in the depletion region of the diode. Anderson and Regan \citep{Anderson_2013} created three-dimensional maps of trap density using an electronic stimulus on the JWST H2RG detectors and showed that traps are not uniformly distributed in the depletion region of the pixels. Since then, many authors worked to characterize the persistence amplitudes and decay as a function of flux, fluence and exposure time. Few phenomenological models exist, some of them describe the persistence decay using an exponential function (\citep{Mosby_2020}, \citep{Tulloch_2019}, \citep{Serra_2015}) while other authors use a power-law (\citep{Long_2012}, \citep{Long_2015}) suggesting that a large range of trapping time constants are responsible for the persistence effect.

In this work, we focus on the persistence of 16 detectors in the NISP focal plane. We show that a power-law model fits better the data than a sum of exponential decays. The amplitudes and persistence decay times are measured and reported and the fluence-dependent model for stimuli below saturation is built to predict persistence levels in NISP images.

\section{PERSISTENCE MODEL FOR NISP DETECTORS}
\label{sec:pers_below_sat}

\subsection{Ground characterisation measurements}
\label{sec:measurements}

We have measured the persistence current as a function of the stimuli amplitudes for all the detectors in the NISP focal plane.
Measurements were done during the detector characterization campaign on ground at the operating temperature of 85 K. 
The stimuli were flat-field illuminations with amplitudes ranging from $5000$ to $95\,000$e. 
For each stimulus level, the persistence current was measured in a dark exposure of 286 frames sampled nondestructively up the ramp UTR(286) following a flat-field exposure of UTR(76) and at each fluence level the measurement was repeated 15 times in a row. 

\subsection{Median persistence amplitudes in the NISP focal plane}
\label{sec:median_persistence}

We first describe the median level of persistence in the 16 detectors after a set of stimuli amplitudes in the range of from $5000$ to $95\,000$e. The integrated persistence charge during the dark exposure is computed for each pixel as the straight line fit to the dark ramp in UTR(276), multiplied by the typical NISP exposure time in photometric exposures. The results are averaged per pixel over the 15 acquired exposures. Then, the median persistence per detector $P_\text{DET}$ and the median persistence from all 16 detectors $P_\text{FPA}$ are computed and used as a reference for the analysis.

As reported in the columns 2 to 4 of Table \ref{tab:PFA_presistence}, after a stimulus of $5000$e, the span of median persistence amplitudes per detector ranges from 1 to 45e with 
\begin{itemize}
    \item 6 detectors having the average persistence lower than 10 electrons, 
    \item 5 detectors having persistence in the interval from 10 to 20 electrons, 
    \item 4 detectors having persistence in the range from 20 to 40 electrons, 
    \item one detector with persistence above 40 electrons. 
\end{itemize}
The persistence contrast $C_P(S)$ between detectors is defined as the relative value of the detector's persistence amplitude with respect to the median persistence amplitude of the focal plane
\begin{equation}
C_P(S) = \frac{P_\text{DET}(S)}{P_\text{FPA}(S)}.
\end{equation}
It ranges from 0.04  to 2.91. The stimulus contrast $C_S = \frac{S_\text{DET}}{S_\text{FPA}}$, where $S_\text{DET}$ is the median signal per detector and $S_\text{FPA}$ is the median signal over the whole FPA, was always below 3\%. Columns 5 to 7 of Table \ref{tab:PFA_presistence} report median persistence values after a stimulus of about $95\,000$ e, close to the pixel full-well level of about $130\,000$e. The span of average persistence amplitudes per detector ranges from about 10 to almost 300e with 
\begin{itemize}
    \item 6 detectors having average persistence lower than 100 electrons, 
    \item 8 detectors having persistence in the interval of from 100 to 160 electrons,  
    \item 2 detectors with persistence higher than 200 electrons. 
\end{itemize}
The contrast in persistence signal ranges from 0.11 for the array with the lower persistence signal up to 2.53 for the array with the highest persistence which is comparable to the contrasts obtained for the low stimulus. In average, the median persistence signal in typical photometric exposures is lower than 1\% of the stimulus for the stimuli below saturation.

\begin{sidewaystable}[ht]
\caption{Median detectors' persistence properties across the NISP focal plane.} 
\label{tab:PFA_presistence}
\begin{center}       
\begin{tabular}{@{}crSScrSScrSScrSS@{}}
\toprule
 & \multicolumn{4}{l}{$S_\text{FPA} = 5\,359$ e} 
 & \multicolumn{4}{l}{$S_\text{FPA} = 94\,634$ e} 
 & \multicolumn{4}{l}{$S_\text{FPA} = 272\,647$ e} 
 & \multicolumn{3}{l}{$S_\text{FPA} = 385\,467$ e} \\
  \addlinespace
 & \multicolumn{4}{l}{$P_\text{FPA} = 15$ e} 
 & \multicolumn{4}{l}{$P_\text{FPA} = 116$ e} 
 & \multicolumn{4}{l}{$P_\text{FPA} = 638$ e} 
 & \multicolumn{3}{l}{$P_\text{FPA} = 746$ e} \\
  \addlinespace
 \textbf{ID}
 &  $P_\text{DET}$ & $\frac{P_\text{DET}}{S_\text{DET}}$ [\%] & $C_P$
 && $P_\text{DET}$ & $\frac{P_\text{DET}}{S_\text{DET}}$ [\%] & $C_P$ 
 && $P_\text{DET}$ & $\frac{P_\text{DET}}{S_\text{DET}}$ [\%] & $C_P$ 
 && $P_\text{DET}$ & $\frac{P_\text{DET}}{S_\text{DET}}$ [\%] & $C_P$ \\
  \addlinespace
\midrule
\textbf{18267} & 45 & 0.88 & 2.91 && 294 & 0.32 & 2.53 && 631 & 0.24 & 0.99 &&  744 & 0.20 & 1.00 \\
\textbf{18632} & 34 & 0.62 & 2.20 && 216 & 0.22 & 1.86 && 621 & 0.22 & 0.97 &&  721 & 0.18 & 0.97 \\
\textbf{18272} & 25 & 0.47 & 1.65 && 159 & 0.17 & 1.37 && 564 & 0.21 & 0.88 &&  639 & 0.17 & 0.86 \\
\textbf{18278} & 25 & 0.46 & 1.61 && 156 & 0.16 & 1.34 && 658 & 0.24 & 1.03 &&  685 & 0.18 & 0.92 \\
\textbf{18284} & 21 & 0.38 & 1.37 && 114 & 0.12 & 0.98 && 362 & 0.13 & 0.57 &&  415 & 0.11 & 0.56 \\
\textbf{18285} & 17 & 0.32 & 1.14 && 121 & 0.13 & 1.04 && 512 & 0.18 & 0.80 &&  586 & 0.15 & 0.79 \\
\textbf{18548} & 17 & 0.31 & 1.10 && 146 & 0.15 & 1.25 && 816 & 0.30 & 1.28 &&  852 & 0.22 & 1.14 \\
\textbf{18453} & 16 & 0.29 & 1.02 && 150 & 0.16 & 1.29 && 792 & 0.29 & 1.24 &&  866 & 0.22 & 1.16 \\
\textbf{18221} & 15 & 0.28 & 0.98 && 113 & 0.12 & 0.97 && 667 & 0.24 & 1.05 &&  819 & 0.21 & 1.10 \\
\textbf{18249} & 15 & 0.28 & 0.97 && 119 & 0.13 & 1.02 && 722 & 0.27 & 1.13 &&  789 & 0.21 & 1.06 \\
\textbf{18628} &  8 & 0.16 & 0.55 &&  93 & 0.10 & 0.80 && 900 & 0.34 & 1.41 && 1020 & 0.27 & 1.37 \\
\textbf{18269} &  7 & 0.14 & 0.48 &&  66 & 0.07 & 0.57 && 227 & 0.08 & 0.36 &&  310 & 0.08 & 0.42 \\
\textbf{18280} &  6 & 0.11 & 0.40 &&  69 & 0.07 & 0.59 && 949 & 0.35 & 1.49 &&  988 & 0.26 & 1.32 \\
\textbf{18268} &  4 & 0.07 & 0.23 &&  29 & 0.03 & 0.25 &&  83 & 0.03 & 0.13 &&   99 & 0.03 & 0.13 \\
\textbf{18452} &  2 & 0.05 & 0.16 &&  25 & 0.03 & 0.22 && 645 & 0.24 & 1.01 &&  824 & 0.22 & 1.10 \\
\textbf{18458} &  1 & 0.01 & 0.04 &&  13 & 0.01 & 0.11 && 443 & 0.17 & 0.69 &&  749 & 0.20 & 1.00 \\
\bottomrule
\end{tabular}
\end{center}
\end{sidewaystable} 


\subsection{Persistence amplitudes - the spatial spread and dynamics per detector}
\label{sec:dynamics}

\begin{figure}[!ht]
\begin{center}
\begin{tabular}{c}
\includegraphics[scale=0.5]{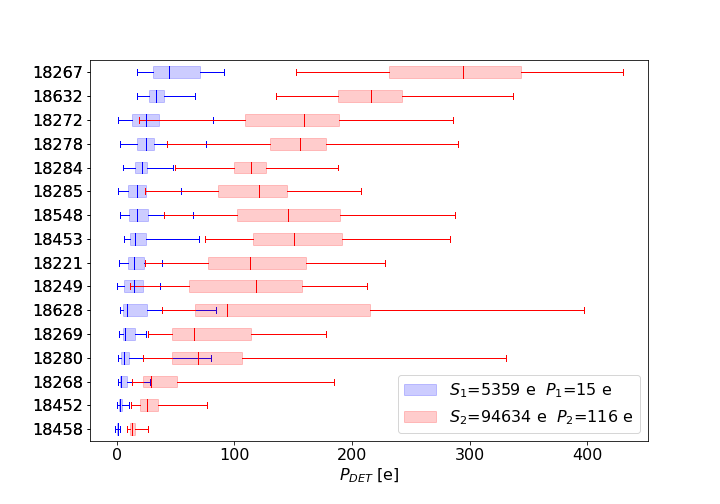}
\end{tabular}
\end{center}
\caption{\label{fig:pers_bxp}Characteristic persistence values per detector after stimuli values of $5000$ (blue) and $95\,000$e (red). The boxes extend from the first to the third quartiles and whiskers extend from the first the 99 percentiles. Inside each box median persistence signal is indicated. The arrays were arranged from the bottom to the top according to the increasing median persistence value at low fluence.}
\end{figure}

\begin{figure}[!ht]
\begin{center}
\begin{tabular}{c}
\includegraphics[scale=0.5]{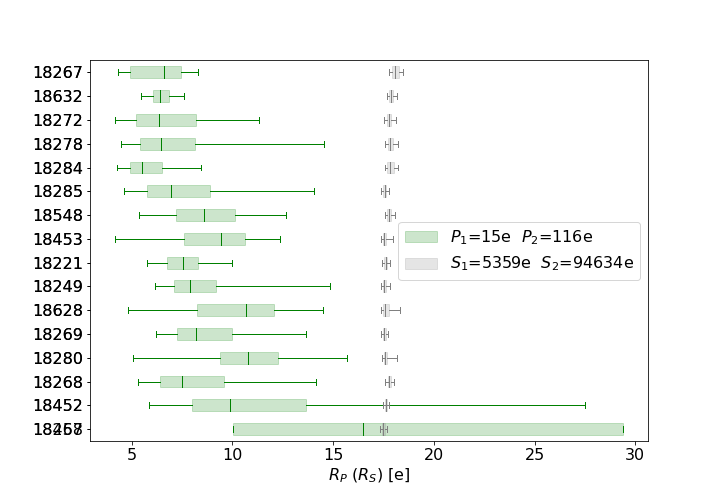}
\end{tabular}
\end{center}
\caption{\label{fig:pers_ratio_bxp}Increase in persistence amplitude $R_P$ per detector (green boxes) compared to the increase in stimuli $R_S$ (grey boxes). The description of boxes is the same as in Fig. \ref{fig:pers_bxp} except the detector 18458 for which the whiskers extend from the 5th to the 95th percentiles for the sake of readability.}
\end{figure}

The persistence signal, and its dependency on the stimulus, is not homogeneously distributed among pixels in single detectors. In the Fig. \ref{fig:pers_bxp} we show the distribution of characteristic persistence values for each detector. Detectors were arranged from bottom to top according to increasing median persistence values. The range of persistence amplitudes covers two (three) orders of magnitude after stimuli of $5000$ ($95\,000$)e.
The increase of persistence amplitude $R_P$, calculated as the ratio of persistence after $S_2=95\,000$e to persistence after $S_1=5000$e stimulus $R_P = \frac{P_\text{DET}(S_2)}{P_\text{DET}(S_1)}$, per detector is shown in the Fig. \ref{fig:pers_ratio_bxp}. For each detector, a factor of almost 20 in the increase in stimulus corresponds to a factor of 5 to 17 in the increase in persistence signal. Interestingly, the detectors with the highest persistence are characterized by the lowest increase in the persistence signal as the stimulus increases. It is quite apparent that persistence is not proportional to the signal.

\subsection{The intra-detector spatial structures of persistence }
\label{sec:structures}

\begin{figure}[!ht]
\begin{center}
\begin{tabular}{c} 
\includegraphics[scale=0.25]{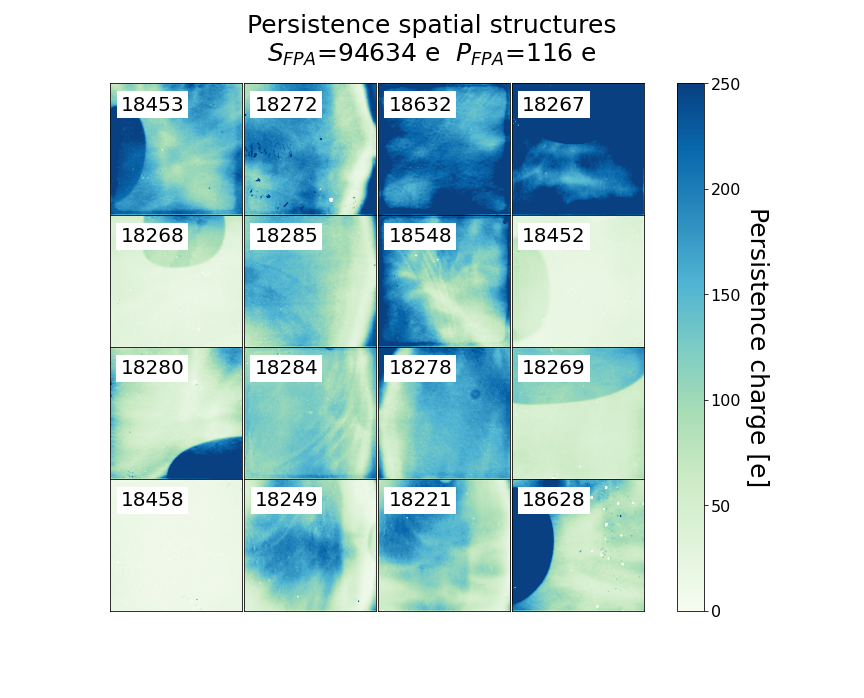}
\includegraphics[scale=0.25
]{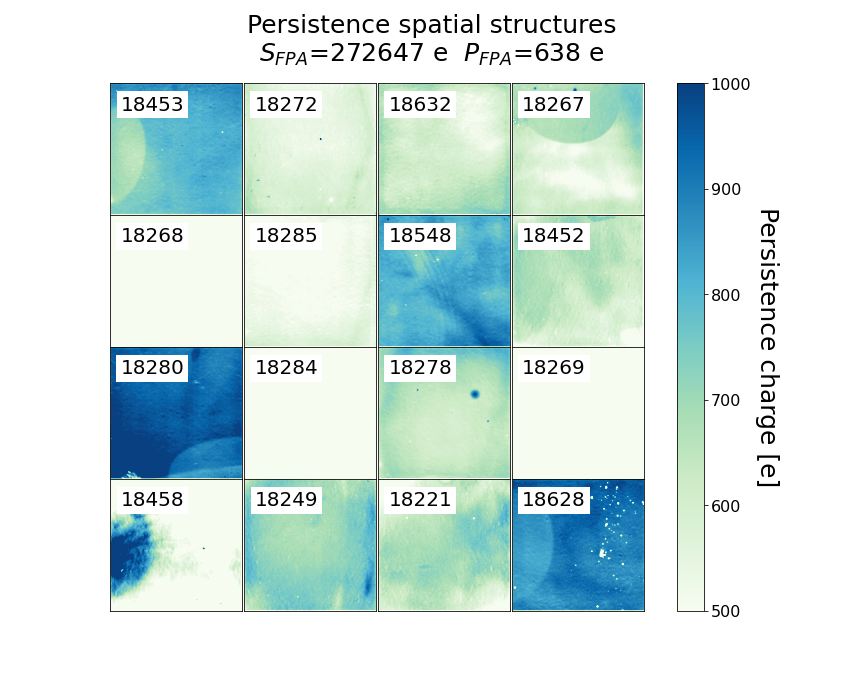}
\end{tabular}
\end{center}
\caption{\label{fig:pers_det_structures}Persistence structures across detectors after a flat-field below saturation (left panel) and above saturation  (right panel).}
\end{figure}

In the left panel of Fig. \ref{fig:pers_det_structures} we show the typical spatial structures that appear in the persistence signal after a flat-field stimuli below saturation. Distinct structures of higher persistence signal with semicircular shapes are visible on more than half of the detectors. On the other six arrays (18272, 18285, 18284, 18632, 18548 and 18278), located in the center of the focal plane, the regions with high persistence are closer to their edges. Two detectors (18249,  and 18221) have irregular zones of high persistence in their centers and only one array seems to be homogeneous (18458). 

We verified that intra-detector regions of low/high persistence, and typical persistence contrast $C_P$ between detectors are preserved for all stimuli below saturation. Also, the ordering of the detectors according to increasing values of the amplitude of the persistence does not change significantly. In particular, the detectors with minimum and maximum persistence amplitudes remain the same, indicating steady type of behaviour of all detectors as the amplitude of the stimulus increases.

\subsection{Persistence model below saturation}
\label{sec:model}

A complete description of the persistence effect requires determining how the initial amplitude of the persistence current varies as a function of the stimulus amplitude and how it decays over time after the stimulus ends. For all fluences analyzed, we verified that the power-law function fits the data better than an exponential or a sum of two or three exponentials, so we adopted this description for the rest of our analysis. The persistent current $I(S,t)$ is thus described as a power-law decay of the form
\begin{equation}
I(S,t) = \alpha(S)\left(\frac{\tau}{t-t_0+\tau}\right)^{\beta(S)},
\label{eq:pers_power_law}
\end{equation}
where $\alpha$ is the initial current amplitude at $t=t_0$, $t_0$ is the time when the stimulus ends and the pixel is reset to its baseline value. The parameter $\tau$ is introduced to avoid divergence at $t=t_0$ and takes into account the difficult-to-measure time delay between the actual end of the illumination and the pixel reset. 

\begin{figure}[!ht]
\begin{center}
\begin{tabular}{cc}
\includegraphics[scale=0.45]{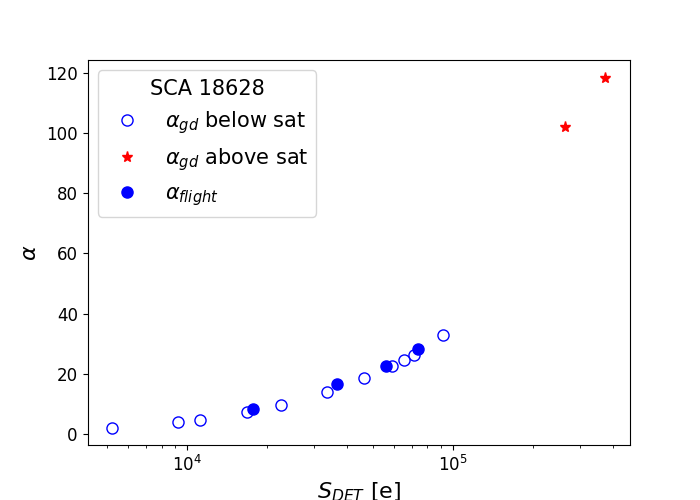}&
\includegraphics[scale=0.45]{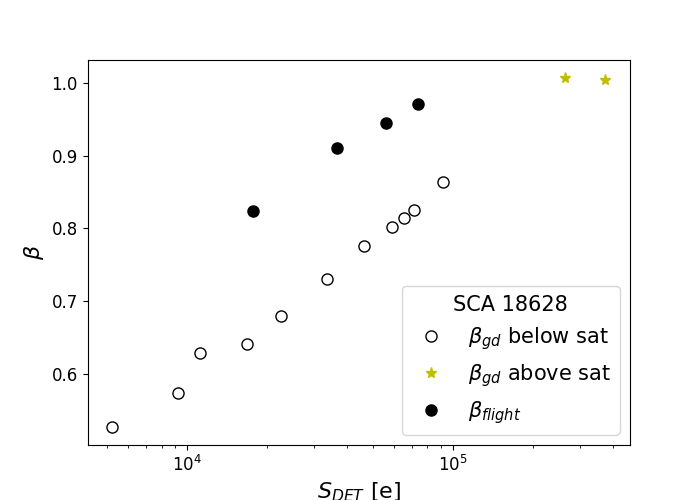}
\end{tabular}
\end{center}
\caption{\label{fig:alpha_beta_median} Median parameters $\alpha$ (left panel) and $\beta$ (right panel) of the persistence model for one detector as function of fluence. The unfilled circles show the parameters $\alpha_\text{gd}$ and $\beta_\text{gd}$ measured during ground characterizations for unsaturating stimuli. The * markers show the parameters $\alpha_\text{gd}$ and $\beta_\text{gd}$ from ground characterizations for saturating stimuli. The filled circular markers represent the coefficients measured from flight calibrations $\alpha_\text{flight}$ and $\beta_\text{flight}$.}
\end{figure}

For each detector, the data were fitted using the integral of Eq. (\ref{eq:pers_power_law}) to estimate $\alpha$ and $\beta$ at each of the analyzed fluences $S$. For each fluence, the $\alpha$ and $\beta$ pixel maps are inferred for each of the 15 dark exposures, and the averaged $\alpha$ and $\beta$ maps are computed accordingly. For illustration purposes, Fig. \ref{fig:alpha_beta_median} shows the evolution of median $\alpha_\text{gd}$ and $\beta_\text{gd}$ parameters as function of stimulus for one detector (the subscript ${}_\text{gd}$ is added to indicate that these are the coefficients calculated using data from ground characterizations).

For the stimuli below saturation the dependence of $\alpha(S)$ can be described as:
\begin{equation}
\alpha(S) =  (a_1+b_1S)\left[1-\exp{\left(-\frac{S}{c_1}\right)}\right],
\label{eq:alpha_fcn_S}
\end{equation}
and the dependence of $\beta(S)$ follows
\begin{equation}
\beta(S) = a_2\left(1 + \frac{S}{b_2}\right)^{c_2}.
\label{eq:beta_fcn_S}
\end{equation}
Hence, the model consists of two steps, first, we fit $\alpha_i$ and $\beta_i$ per fluence $S_i$ on exposures up the ramp, and then we fit the dependence of $\alpha$ and $\beta$ on the stimulus amplitude $S$ for all pixels independently. The final model, therefore, has 6 parameters for each pixel ($a_1$, $b_1$, $c_1$ to describe the dependence of $\alpha(S)$ and $a_2$, $b_2$, $c_2$ to describe the dependence of $\beta(S)$). 

\begin{figure}[!ht]
\begin{center}
\begin{tabular}{c} 
\includegraphics[scale=0.45]{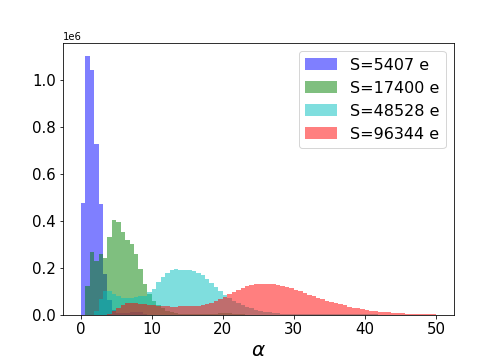}
\includegraphics[scale=0.45]{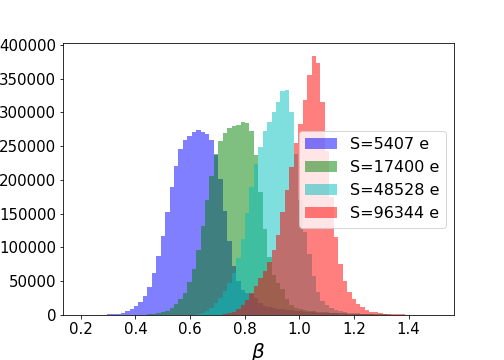}
\end{tabular}
\end{center}
\caption{\label{fig:alpha_beta_disrtb}Distribution of persistence model parameters $\alpha$ and $\beta$ calculated using Eqs. (\ref{eq:alpha_fcn_S}) and (\ref{eq:beta_fcn_S}) knowing the fluence $S$ for one detector.}
\end{figure}

\begin{figure}[!ht]
\begin{center}
\begin{tabular}{c}
\includegraphics[scale=0.45]{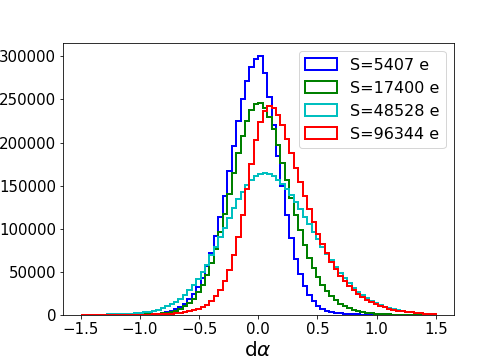}
\includegraphics[scale=0.45]{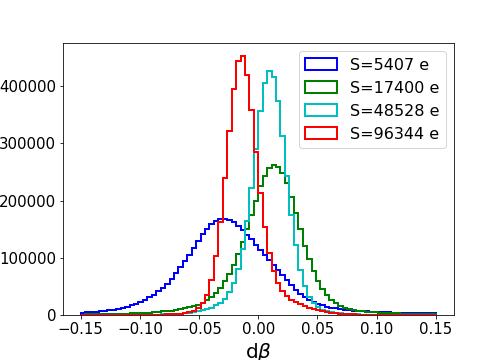}
\end{tabular}
\end{center}
\caption{\label{fig:alpha_beta_errors}Difference between the $\alpha$ and $\beta$ measured per fluence and the model prediction using Eqs. (\ref{eq:alpha_fcn_S})  and (\ref{eq:beta_fcn_S}) for one detector.}
\end{figure}

An example of the spread of the model parameters $\alpha$ and $\beta$ calculated using Eqs. (\ref{eq:alpha_fcn_S})  and (\ref{eq:beta_fcn_S}) knowing the fluence $S$ per pixel and their evolution with $S$ is shown in Fig. \ref{fig:alpha_beta_disrtb} for one detector. In Fig. \ref{fig:alpha_beta_errors} we show the difference between the averaged $\alpha$ and $\beta$ measured per fluence and the model prediction shown in Fig. \ref{fig:alpha_beta_disrtb}. 

\subsection{Precision of the persistence model from ground characterisations}
\label{sec:precision}

The main drivers of our analysis are the model's precision and accuracy in predicting the persistence charge, as these determine the model's ability to effectively mask or correct the persistence signal during routine observations of the \Euclid telescope.

\begin{figure}[!ht]
\begin{center}
\begin{tabular}{cc}
\includegraphics[scale=0.25]{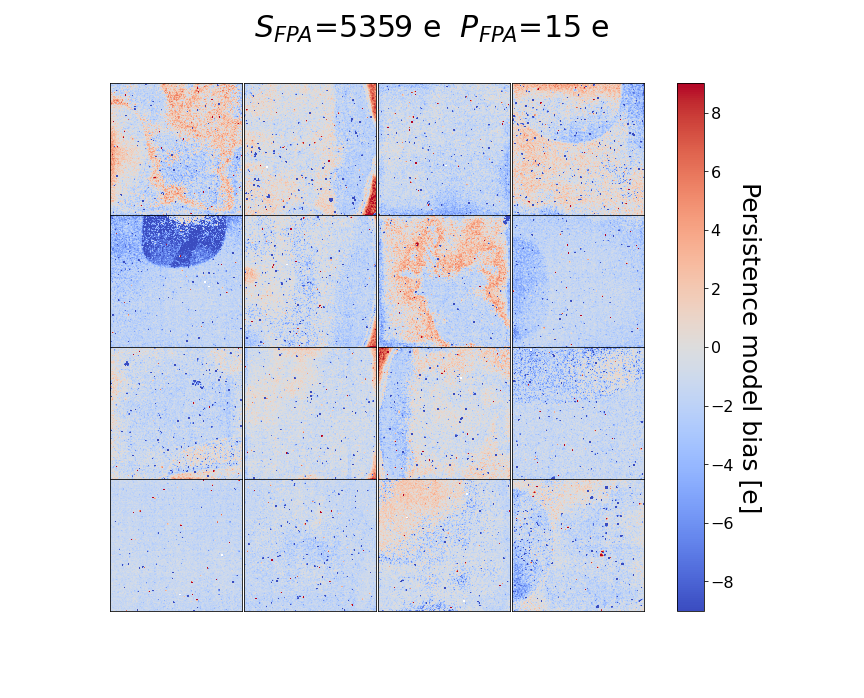} &
\includegraphics[scale=0.25]{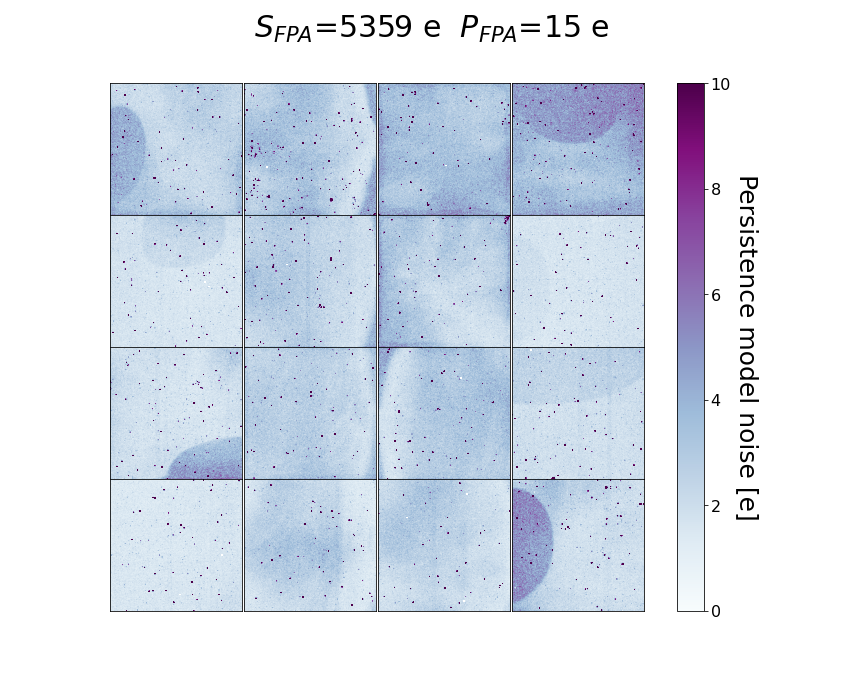} \\
\includegraphics[scale=0.25]{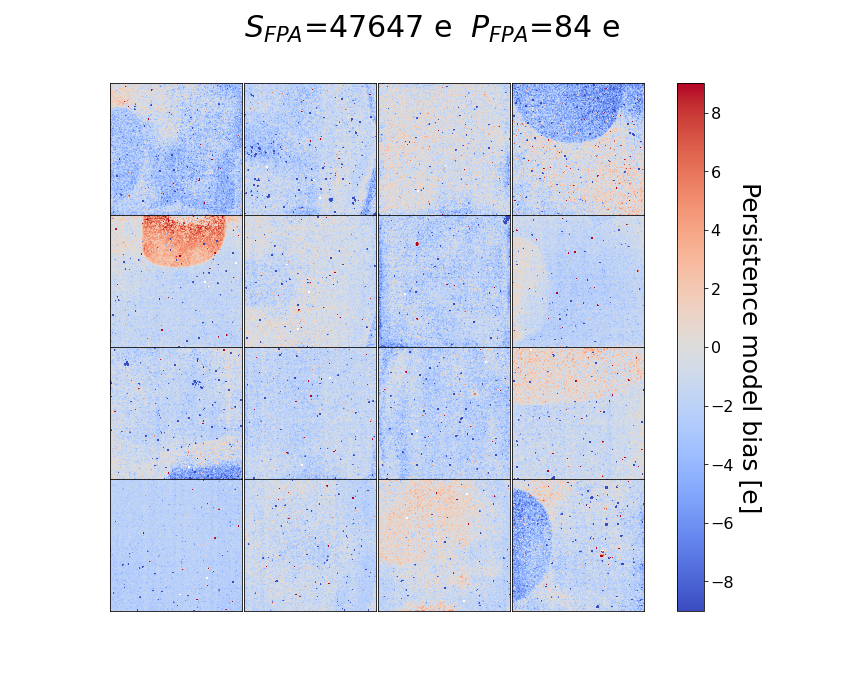} & 
\includegraphics[scale=0.25]{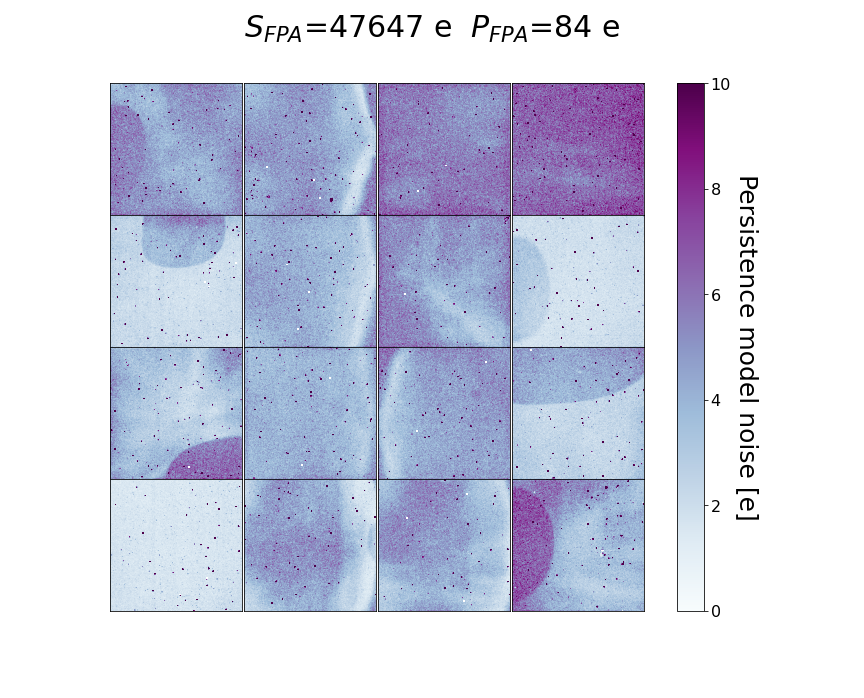} \\
\includegraphics[scale=0.25]{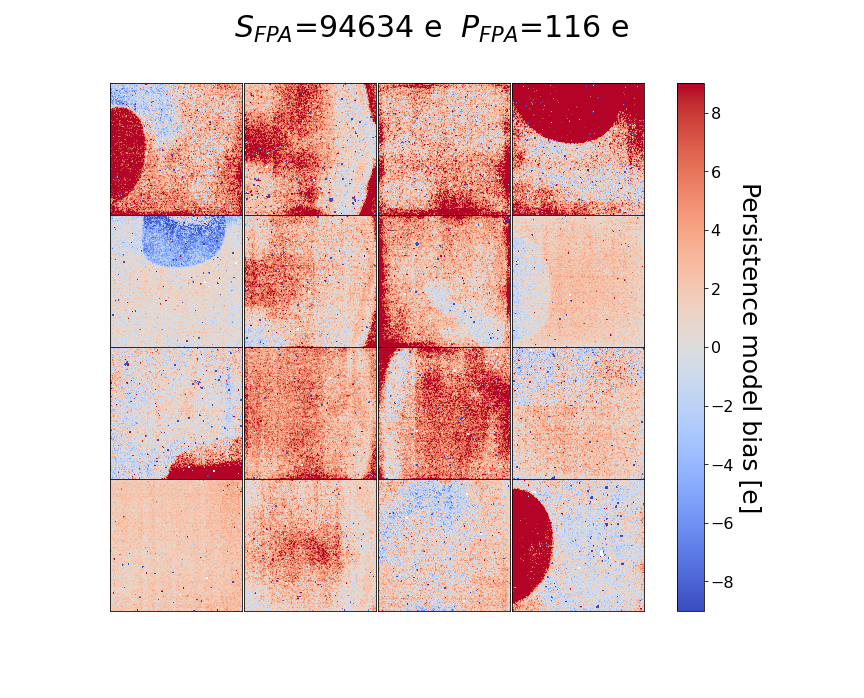} &
\includegraphics[scale=0.25]{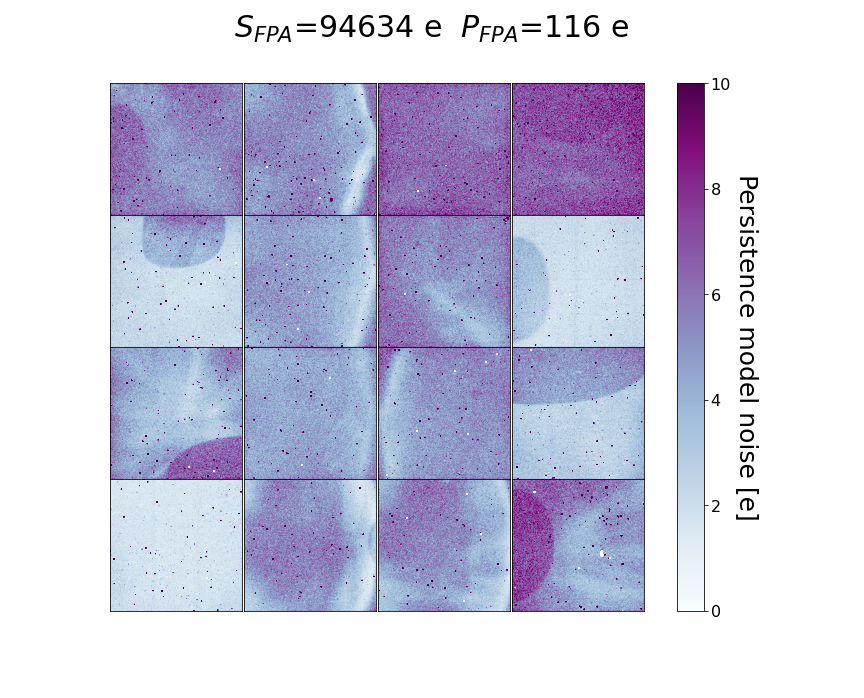}
\end{tabular}
\end{center}
\caption{\label{fig:power_law_model_pixel_accuracy}Spatial distribution of the bias (left column) and of the noise (right column) of the power-law model after 3 fluences below saturation (from top to bottom: bout $5000$, $50\,000$ and $95\,000$e). }
\end{figure}

The accuracy (bias) of the signal-dependent model was calculated per pixel as the average (over 15 sequences for the same fluence) difference between the linear slope fitted on the model ramp in MACC($n_\text{g}$,$n_\text{f}$,$n_\text{d}$)\footnote{The NISP detectors acquire data using a so-called Multi-Accumulation MACC($n_{\text{g}}$,$n_{\text{f}}$,$n_{\text{d}}$) sampling in which $n_{\text{g}}$ groups of $n_{\text{f}}$ averaged consecutive frames are separated by $n_{\text{d}}$ dropped frames. (See \citep{Kubik:2016hsg} and \citep{NISP_prc})} generated using Eqs. (\ref{eq:pers_power_law}),  (\ref{eq:alpha_fcn_S}) and (\ref{eq:beta_fcn_S}) and the linear slope fitted on the data in the same MACC mode. 
Spatial distribution of bias is shown in the left column of the Fig. \ref{fig:power_law_model_pixel_accuracy} for three different fluences. The model tends to overestimate the persistence by few electrons for low stimuli, while for stimuli close to saturation there is a little underestimation of the persistence signal. The bias ranges from about 5\% to about 10\% of the persistence signal and is a major challenge for the model.

The model noise, measured as per pixel standard deviation over 15 sequences of residuals at each fluence level, is shown in the right column of Fig. \ref{fig:power_law_model_pixel_accuracy}. The noise scales with increasing fluence, which is expected due to the shot noise contribution. The median noise per detector is lower than 8e r.m.s. for all stimuli.


\subsection{Verification of the persistence amplitudes and decays in flight}

Special acquisitions to measure persistence were taken during Performance Verification (PV) phase of the mission. The acquisition strategy consists of flat-field exposures at four different flux intensities (below the saturation), each followed by a sequence of 15 dark exposures. The first five dark exposures are acquired in photometric mode MACC(4,16,4) -- with an integration time of about 87s, while the remaining ten exposures in spectroscopic mode MACC(15,16,11) -- with an integration time of about 550s. This strategy allows the measurement of persistence charge decay over more than 1.5 hours. An analysis similar to that performed on data from the ground characterisation was performed on PV data. The difference was that for ground characterisation data we had access to all the frames acquired up the ramp, while for flight data we only had access to the flux integrated over 87 or 550s. The onboard processing of NISP data averages out the contribution of the persistence current, which -- as inferred from the ground data -- varies significantly over these time ranges. This reduces the precision of the model for short-term persistence decay, but allows estimating the decay rate on longer time scales. The latter is very important for the NISP instrument, which continuously scans the sky and afterimages of bright sources can interfere with the signal from other observations over long time intervals. We have fitted the model parameters from flight measurements (shown as filled circular markers in Fig. \ref{fig:alpha_beta_median} for one detector) to be on average compatible with the ones estimated over short exposures of 10 minutes from ground characterisations. Trends characteristic of coefficients obtained from ground calibrations are also present in the flight data. The $\alpha_{\text{flight}}$ coefficients fall within the same range for flight and ground measurements and, as expected, increase with fluence. The increase in $\beta_{\text{flight}}$ parameters with fluence is still present, confirming our measurements during ground calibrations. The physical interpretation of the differences between ground and flight data, which can be significant for some detectors, is not obvious, because not only did the operating conditions of the NISP instrument changed, but also the way the raw NISP data were processed and reduced.

\subsection{Discussion of the results}
\label{sec:discussion}

The initial persistence current amplitude $\alpha$ increases nonlinearly with stimulus $S$. The increase in $\alpha$ is usually explained by the increase of the number of traps exposed to carriers as the edge of the depletion region moves towards the PN junction (\citep{Smith_2008a}, \citep{Long_2013}). The nonlinearity of the increase, in turn, can be interpreted as the lower density of trapping centers close to the edge of the PN junction assuming a uniform distribution of the trapping and release time constants in the bulk or alternatively, if both the density and time constants are homogeneously distributed, then we can attribute it to the nonlinear increase of the depletion region with fluence. Most likely, both effects can be present and affect the shape of the signal dependence of the persistence amplitude. 

The fact that $\beta$ depends on the stimulus amplitude is more challenging to explain. We cannot interpret this trend as being caused by the superposition of persistence signals coming from several sources since the superposition effect results in a reduction of the power-law index. For unsaturated stimuli, the authors of \citep{Smith_2008b} report that the persistence decay shape is invariant with stimulus amplitude, suggesting that the decay time constants are proportional to the exposure time. They also report that the persistence decay rate depends on both the duration and intensity of the stimulus. On the other hand, Long in \citep{Long_2013} notes that the measured slopes at low fluence levels are steeper than at higher fluence levels, and attributes this to either the short period of time the traps are exposed to free carriers or the fact that traps exposed to lower fluence levels may have shorter release times. The increase of the power-law index with stimulus flux observed in our analysis suggests steeper persistence decay after higher fluence. Since the exposure time was the same for all stimulus intensities, this may indicate fast traps located closer to the PN junction and that traps with longer trapping times have longer release times as already suggested in \citep{Long_2013}. It is also compatible with the statement that the trapping and release time scales are comparable to the exposure times (\citep{Long_2013}, \citep{Long_2015}). The increase of $\beta$ for high persistence detectors is striking. We note that $\beta \leq 1 $ implies an infinite charge as $t \to\infty$. Hence, the parametrisation can only be valid in a finite time interval.

\section{Trapping stabilisation and superposition}
\label{sec:pers_stab}

To measure the time needed for a detector to reach a steady-state of capture and release, we placed the detectors in the dark conditions for a long period of time before starting the persistence measurements using our nominal measurement scheme of 15 pairs of flat and dark exposures at constant nominal fluence. 

\begin{figure}[!ht]
\begin{center}
\begin{tabular}{cc} 
\includegraphics[scale=0.4]{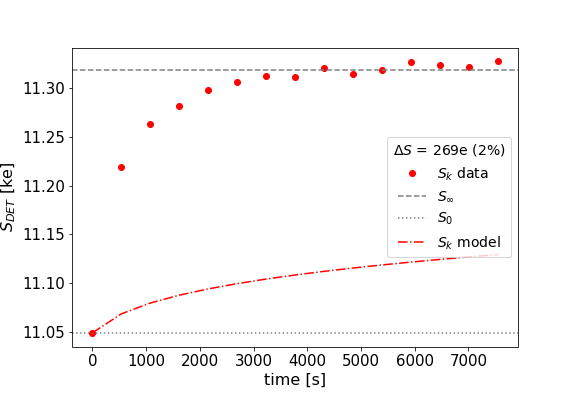}&
\includegraphics[scale=0.4]{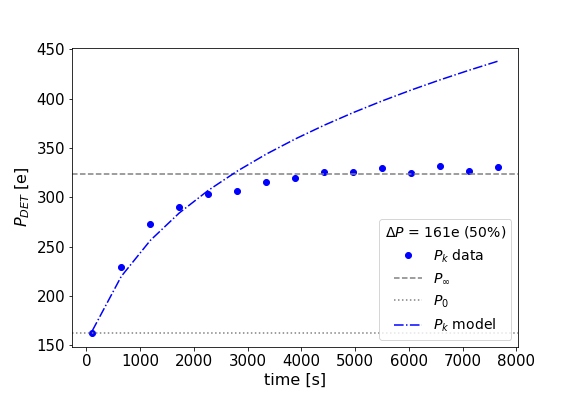}
\end{tabular}
\end{center}
\caption{\label{fig:pers_stabilisation}Stabilisation of detector median stimulus amplitude (left panel) and persistence amplitude (right panel) in a sequence of measurements consisting of 15 pairs of flat and dark exposures at a constant nominal fluence, taken after the detector has been in the dark for a long period of time. The steady-state values of signal $S_\infty$ and persistence $P_\infty$ amplitudes are indicated by grey dashed lines.}
\end{figure}

The measurement results for one of the detectors are shown in the Fig. \ref{fig:pers_stabilisation}. The red (blue) circles in the left (right) panel show the measured median stimulus (persistence) amplitude evolution over time. The steady-state values $S_\infty$ and $P_\infty$, computed as the average over the last 10 measured values, are indicated by horizontal dashed lines. We observe that both the median amplitude of the stimulus and the median amplitude of the persistence take about 30 minutes to reach a steady-state for the detector under study and that the increase in stimulus and persistence amplitudes between the first measurement in the sequence, hereafter $S_0$ and $P_0$ indicated as grey dotted lines in Fig.\ref{fig:pers_stabilisation} , and the steady-state values is of about 270 and 160e respectively. This corresponds to an increase $\Delta S = S_\infty - S_0$ of 2\% in case of stimulus amplitude  and an increase  $\Delta P = P_\infty - P_0$ of 50\% in the case of persistence amplitude.

We also check whether the increase in stimulus and persistence amplitudes are due to the accumulation and superposition of persistence currents from previous exposures. 
To determine this, we compare the data with the predictions of the model, in which we assume that the $k$-th measurement in the sequence was affected by the additive contribution of persistence currents. Namely, for stimuli amplitudes we assume that:
\begin{equation}
S_k = S_0 + \sum_{j=0}^{k-1}\int_{t_\text{i}}^{t_\text{e}}\alpha(S_j)\left(\frac{\tau}{t-t_{0,j}+\tau}\right)^{\beta(S_j)}dt,
\end{equation}
and for persistence amplitudes we assume that:
\begin{equation}
P_k = \sum_{j=0}^{k}\int_{t_\text{i}}^{t_\text{e}}\alpha(S_j)\left(\frac{\tau}{t-t_{0,j}+\tau}\right)^{\beta(S_j)}dt,
\end{equation}
where $t_\text{i}$ and $t_\text{e}$ are the begin time and end time of exposure $S_k$ and $P_k$ while $t_{0,j}$ is the end of exposure $S_j$. The results are shown in Fig. \ref{fig:pers_stabilisation} as red and blue dash-dotted lines respectively. 
Clearly, the increase in the stimulus amplitude cannot be explained only by the accumulation of persistence currents, its contribution is too small. The apparent inefficiency of signal detection can be assigned to trapping. After the detector has been in the dark for a long time, we can assume that all the traps have been freed and during the first measurement of the stimulus they are filled causing a drop in the effective efficiency of signal detection. Thus, we are tempted to interpret the difference between the data points and the model prediction as the trapping of charges not taken into account in our model. A long time needed to reach a steady-state relative to the detrapping profiles indicates the  the time constants of capture are much shorter (comparable or shorter than exposure time) than time constants of release (comparable or longer than exposure time), in agreement with suggestions of \citep{Tulloch_2019} while \citep{Long_2013} proposes that trapping times could be comparable or longer than exposure times.

As for the increase in persistence amplitude, the model's predictions are in good agreement with the data points until the detector reaches steady-state. The power-law index decreases over time for the analyzed sequence of measurements, indicating a milder variation of the persistence current. This is consistent with the assumption of superposition of persistence currents over time. This means that we can indeed describe persistence as a superposition of currents coming from several previous sources in a situation when the detector has not been exposed to strong and regular stimuli for a long time, which is often the case during nominal observations of the \Euclid telescope.

\section{Persistence above saturation}
\label{sec:pers_above_sat}

In addition to the stimuli below saturation described in section \ref{sec:pers_below_sat}, for the same operating conditions of NISP ($T \sim 85$ K), during the ground tests we measured the persistence decay for two fluences above saturation, namely $270\,000$ and $380\,000$e, representing almost four times the full-well capacity of NISP detectors.  The persistence current was measured in a dark exposure in UTR(286) following a flat-field exposure of UTR(394). This procedure was repeated 15 times for each of the stimulus.

\subsection{Persistence amplitudes and model parameters for stimuli above saturation}

The median persistence amplitudes $P_\text{FPA}$ after stimuli above saturation reach levels of about 600 and 700e as reported in last six columns of Table \ref{tab:PFA_presistence}. The spatial pattern of low/high persistence detectors changes with respect to what was observed for signals below saturation, and the persistence contrast distribution becomes narrower with the lowest value of 0.17 reaching a maximum around 1.5.  

\begin{figure}[!ht]
\begin{center}
\begin{tabular}{c} 
\includegraphics[scale=0.5]{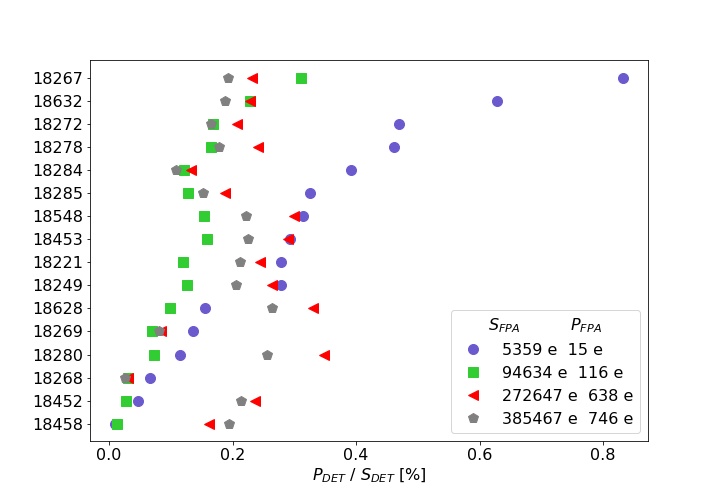}
\end{tabular}
\end{center}
\caption{\label{fig:pers_to_signal}Relative persistence in NISP detectors for typical photometric exposure of 87s. Unsaturating stimuli are represented by blue circles and green squares. Saturating fluences are represented by red triangles and grey pentagons.}
\end{figure}

The comparison of the relative persistence signal obtained for two unsaturating and two saturating stimuli is shown Fig. \ref{fig:pers_to_signal}. For detectors in the upper part of the diagram (top six arrays) we observe that the relative persistence was very high for lowest fluence of $5000$e (blue circles), while for stimuli close to (green squares) and above saturation (red triangles and grey pentagons) the median increase in persistence is almost linear and does not depend on fluence amplitude. A similar behavior is observed for two arrays in the bottom part of the chart: 18268 and 18269. For the detectors in the central and bottom part median relative persistence decreases for fluences below saturation while for saturating stimuli we see again a fast nonlinear increase in persistence values, so that the median relative persistence for saturating stimuli is higher than for fluences below saturation.  The model parameters $\alpha$ and $\beta$ do not follow the same trend as measured for unsaturating stimuli, as can be seen in Fig. \ref{fig:alpha_beta_median} (* markers). The power-law index is for most of the detectors higher if the fluence is above saturation, which indicates steeper decay. Only three detectors, with low persistence amplitudes, behave differently. 


\subsection{Persistence pattern inversion}

If we consider the persistence spatial structures inside each of the detectors, we note a reversal of areas of high and low persistence amplitudes compared to what is observed for fluences below saturation. In the Fig. \ref{fig:pers_det_structures} we show the persistence structures after stimuli close to but below saturation (left panel) and a stimuli above saturation (right panel). The detectors identified in the central and bottom part of the Fig. \ref{fig:pers_to_signal} and characterized by a rapid increase in persistence values, have indeed large areas where the initially lower persistence signal relative to the rest of the detector pixels has now become relatively much higher. This suggests that at saturation other kind of high-density traps become exposed, which are probably located near the PN junction or at the interfaces. An inversion of the persistence pattern can also be observed in some detectors with a linear increase in persistence for stimuli above saturation, identified in the top part of Fig. \ref{fig:pers_to_signal} but the regions which become high-persistence are smaller in size.


\subsection{Long-term alteration of state under the influence of saturating stimuli}
\label{sec:pers_alter}

Since persistence after stimulus saturation changes its state, we made measurements to see if this change remains over time, or if the detector quickly returns to the initial state observed before saturation. The nominal measurement scheme of 15 pairs of flat and dark exposures at constant  fluence was repeated twice for the same nominal fluence well below saturation. The first sequence of measurements was made before the delivery of any saturating stimulus. A saturating stimulus was then delivered, after which the detector was placed in the dark conditions during 6 hours allowing the persistence current time to decay, followed by a second sequence of measurements. 

\begin{figure}[!ht]
\begin{center}
\begin{tabular}{cc}
\includegraphics[scale=0.45]{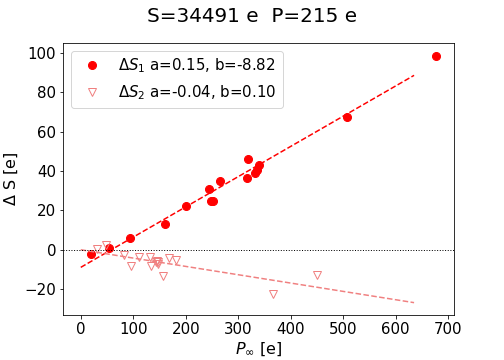}&
\includegraphics[scale=0.45]{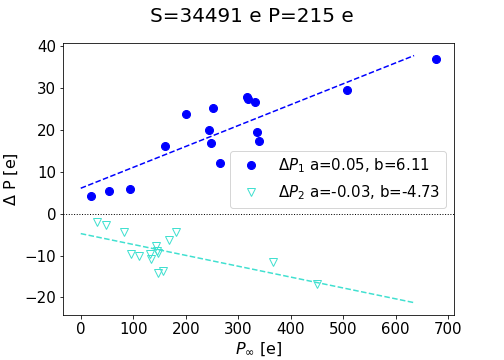}
\end{tabular}
\end{center}
\caption{\label{fig:DS12_vs_P} Dependence of $\Delta S = S_\infty - S_0$ (left panel) and $\Delta P = P_\infty - P_0$ (right panel) on the persistence amplitude $P_\infty$. The circular markers represent measurements taken before the saturating stimulus was delivered. The unfilled triangular markers represent measurements taken after the delivery of a saturating stimulus.}
\end{figure}

A comparison of measurements taken before and after the saturation signal was delivered shows that the detector has not returned to the initial state. The stimulus amplitudes $S_{0,1}$ and $S_{\infty,1}$ measured before saturation are in general lower than the initial amplitudes $S_{0,2}$ and $S_{\infty,2}$ measured after saturation ($S_{\infty,1} < S_{\infty,2}$). Whereby the difference between the steady-state and the initial amplitudes of signal $\Delta S_{1/2} = S_{\infty,1/2} - S_{0,1/2}$ is in both cases proportional to the persistence amplitude, but has the opposite sign as shown in the left panel of Fig. \ref{fig:DS12_vs_P}. $\Delta S_2 < 0$ means that the measured signal amplitude decreases over time. 

Similarly, we observe a change in the behavior of the persistence current before and after the delivery of a saturating stimulus. While before saturation, the amplitude of persistence increased during the sequence of successive measurements, due to the superposition of signals, after saturation the persistence currents decrease over time ($\Delta P_2 = P_{\infty,2} - P_{0,2} < 0$ as shown in the right panel of Fig. \ref{fig:DS12_vs_P}. 
The amplitudes of persistence measured before saturation are in general higher than after the delivery of a saturating stimulus  ($P_{\infty,1} > P_{\infty,2}$).

Negative values of $\Delta S_2$ and $\Delta P_2$ indicate a long relaxation time after saturating stimulus. The increase in the detected signal $S_{\infty,1} < S_{\infty,2}$ and the decrease in persistence charge $P_{\infty,1} > P_{\infty,2}$ after a delivery of a saturating stimulus might be understood as less trapping and less release after saturation as if some of the traps have been deactivated by saturation.

\section{Summary and conclusions}
\label{sec:summary}

During the ground characterisation campaign of NISP detectors' persistence signal was measured for stimuli below and above saturation. The persistence amplitudes are on the order of 1\% of the stimulus and vary significantly from detector to detector. The measurements of persistence during flight calibrations show no significant change in persistence with respect to ground characterizations  but the direct comparison is challenging due to different measurement scheme and different data reduction steps. The signal-dependent persistence model derived in this work for stimuli below saturation that describes persistence signals with the accuracy of a few electrons. In this work, we also show the persistence signal after a stimulus above saturation, as well as the durable change in detector response over time after stimulation with a saturating stimulus.

\begin{acknowledgements}

\AckEC
\end{acknowledgements}

\bibliography{paper} 
\bibliographystyle{plainnat} 

\end{document}

%% file: authors.tex
\author[1]{On behalf of the Euclid Collaboration: B.~Kubik\thanks{\hspace{0.2cm}\url{bkubik@inpl.in2p3.fr}}\hspace{0.1cm}}
\author[1]{R.~Barbier}
\author[1]{G.~Smadja}
\author[1]{S.~Ferriol}
\author[1]{Y.~Conseil}
\author[1]{Y.~Copin}
\author[6]{W.~Gillard}
\author[2]{S.~Dusini}
\author[3]{K.~Jahnke}
\author[4]{E.~Prieto}
% ...

\affil[1]{{\small Universit\'e Claude Bernard Lyon 1, CNRS/IN2P3, IP2I Lyon, UMR 5822, Villeurbanne, F-69100, France}\vspace{-1mm}}
\affil[2]{{\small INFN-Padova, Via Marzolo 8, 35131 Padova, Italy}\vspace{-1mm}}
\affil[3]{{\small Max-Planck-Institut f\"ur Astronomie, K\"onigstuhl 17, 69117 Heidelberg, Germany}\vspace{-1mm}}
\affil[4]{{\small Aix-Marseille Universit\'e, CNRS, CNES, LAM, Marseille, France}\vspace{-1mm}}
% ...
\affil[6]{{\small Aix-Marseille Universit\'e, CNRS/IN2P3, CPPM, Marseille, France}\vspace{-1mm}}
% ...
